\documentclass[aps,prd,onecolumn,groupedaddress,showpacs,nofootinbib,amssymb]{revtex4}
\usepackage{graphicx,bm,color}
\usepackage{amsmath}
\usepackage{amssymb}
\usepackage{amsfonts}
\usepackage{comment}

\newcommand{\be}{\begin{equation}}
\newcommand{\ee}{\end{equation}}
\newcommand{\bea}{\begin{eqnarray}}
\newcommand{\eea}{\end{eqnarray}}
\newcommand{\beaa}{\begin{eqnarray*}}
\newcommand{\eeaa}{\end{eqnarray*}}

\newcommand{\nn}{\nonumber \\}
\newcommand{\e}{\mathrm{e}}

\allowdisplaybreaks[4]

\begin{document}

\title{Self-interacting charged massive spin two particles in Minkowski spacetime}

\author{Yuichi Ohara}

\affiliation{
Department of Physics, Nagoya University, Nagoya 464-8602, Japan \\
%$^2$ Kobayashi-Maskawa Institute for the Origin of Particles and
%the Universe, Nagoya University, Nagoya 464-8602, Japan 
}

\begin{abstract}
A model of the self-interacting charged massive spin-two field is constructed. We investigate several properties of the 
model and find that the trivial vacuum is only allowed due to the internal symmetry. This suggests that 
the Higgs mechanism might not be induced by the model of the massive spin-two field with the ghost-free potential.

\end{abstract}

\pacs{95.36.+x, 12.10.-g, 11.10.Ef}

\maketitle

%ゴーストフリー構造のアドバンテージはスピン2の場の理論の場合2つある。1. カットオフスケールが大きくなる。2.非自明な真空の元で（任意のバックグラウンド）で
%ゴーストがでない。

\section{Introduction \label{Sec1}}
Theories of  massive gravity have made a rapid progress over the past decade inspired by the discovery of the late time acceleration.
Although the free field theory proposed by Fierz and Pauli was formulated about 75 years ago \cite{Fierz:1939ix}, the gravitational 
theory describing interaction between massive spin-two particles had not been established until recently because of a ghost problem.
Naively, the interaction terms for massive spin-two particles seem arbitrary due to the presence of the mass term in analogy with the theory of the Proca field. 
Unfortunately, the story does not hold in the case of massive spin-two particles and Boulware and Deser showed that interaction terms generally induce a ghost 
\cite{Boulware:1974sr}. This fact is called the Boulware-Deser ghost problem and had prevented construction of a theory of massive gravity.
The breakthrough came from series of papers about the late time acceleration. In the early 2000s, 
the Dvali-Gabadadze-Porrati brane world model (DGP model) \cite{Dvali:2000} and massive gravity attracted much 
interest for the explanation of the tiny value of the cosmological constant. A field theoretical analysis of the DGP model 
and massive gravity \cite{Luty:2003vm, Nicolis:2004qq, ArkaniHamed:2002sp} gave an important clue to the resolution of the ghost problem and 
de Rham, Gabadadze and Tolley formulated the first ghost-free massive gravity called the dRGT massive gravity \cite{deRham:2010ik, Hassan:2011hr, Hassan:2011ea, Hassan:2011tf, Hassan:2011vm}.
The extended theory containing two dynamical metrics was also formulated and Hassan and Rosen showed that the theory is really ghost-free \cite{Hassan:2011ea,Hassan:2011zd}.

After the formulation of the first ghost-free nonlinear massive gravity, there are several works on constructing new terms
toward the generalization of the dRGT theory. Hinterbichler attempted to give a new kind of interaction terms for the theory and 
discovered new derivative interactions for the Fierz-Pauli theory \cite{Hinterbichler:2013eza} and conjectured the existence 
of the nonlinear counterpart. Unfortunately, these derivative 
interaction terms turned out to have no nonlinear counterpart \cite{deRham:2013tfa} but, it was shown in the Hamiltonian analysis that 
the leading term of the potential in the dRGT theory keeps the degrees of freedom of the massive spin-two field.
We focused on this fact and constructed the new massive spin-two model consisting of the linearized Einstein-Hilbert term and 
the finite potential terms \cite{Ohara:2014vua}. The leading terms of the potential in the dRGT theory ensure that the theory is also ghost-free around nontrivial vacua and we investigated the property 
of the theory around each vacua \cite{Ohara:2015vua}. The motivation for this model is to ask if the massive spin-two field theory 
has to be regarded as a modification of gravity. In many cases, we start the construction of a massive spin-two model under the assumption that the kinetic term should be the Einstein Hilbert 
term but there is no reason why we believe this assumption. 
%In the case of massless spin-two particles, Lorentz invariance and quantum mechanics require the gauge symmetry and the invariance of the S matrix under the gauge transformation leads to the universality of the strength of couplings. This fact forces the theory of massless spin two fields 
%to have the Einstein-Hilbert term as a kinetic term.  On the other hand, the universality of the couplings is not required in the case of the massive theory. This means that there is no 
%need to assume the Einsten Hilbert term in priciple. 
To clarify the difference between the dRGT massive gravity and the new model we proposed, we also considered our model in a curved spacetime by assuming that the spin-two field is not a perturbation of a background 
metric and found that the new theory is consistent only if the background spacetime has the maximal symmetry \cite{Akagi:2014dec} as in the case of the Fierz-Pauli theory 
in a curved spacetime \cite{Buchbinder:1999ar,Buchbinder:1999be}. Furthermore, we derived the general interactions allowed in the Einstein manifold.

There are some previous works which do not regard the theory of massive spin-two particles as the theory of gravity. 
While some discussed the consistency of the massive spin-two field as an alternative gravity theory from the late 1950s to the mid 1970s,
Federbush worked on construction of a field theoretical model describing the dynamics of the charged massive spin-two 
particles. Federbush constructed the $U(1)$ invariant Fierz Pauli action and replaced partial derivatives with covariant derivatives to
introduce the $U(1)$ gauge field into the theory \cite{Federbush}. 
His study revealed that the noncommutativity of the covariant derivatives gives an ambiguity to the definition of the kinetic term 
but the requirement of the correct number of the degree of freedom  uniquely determines the theory. On the other hand, it is well known 
that the theory exhibits acausality for arbitrary values of the background electromagnetic field \cite{Kobayashi, Shamaly, Deser}.

Since the dRGT massive gravity is considered as the general action containing all interaction terms between neutral spin-two particles, it is 
expected that the more general charged spin-two action can be obtained by modification of the dRGT massive gravity action.
de Rham, Matas, Ondo and Tolley considered this kind of extension in \cite{de Rham:2014ci}, but they showed algebraically that the Einstein-Hilbert action is 
not compatible with $U(1)$ symmetry and the Einstein-Hilbert term should be modified. Unfortunately, according to \cite{deRham:2013tfa}, the modification 
necessarily leads to the undesirable ghost mode. Therefore, we cannot write down the $U(1)$ invariant massive gravity action. On the other hand, 
our model proposed in the previous works consists of the linearized kinetic term and interaction terms only. 
This suggests that we could potentially construct the $U(1)$ invariant massive spin-two action by extending the model in \cite{Ohara:2014vua}.

In this paper, we build the self-interacting charged model using the quartic interaction proposed by Hinterbichler and have shown that the theory is really ghost-free. 
Furthermore, we study the behavior of the massive spin-two field around the vacua which stem from the potential term and we reveal the 
parameter region of the theory where the particle description could hold. As a result, we find the $U(1)$ charge puts on the additional constraint on
the parameter space and, as a result, the trivial vacuum is uniquely chosen by this internal symmetry.

\section{New model of massive spin-two particle}

In \cite{Ohara:2014vua}, we construct the new theory of the massive spin-two particle with interaction.
The Lagrangian of the free massive spin-two particle consists of the linearized Einstein-Hilbert action 
and the Fierz-Pauli mass term \cite{Fierz:1939ix},
\be
\label{FPLag}
\mathcal{L}_{\mathrm{FP}} = -\frac{1}{2}\partial_\lambda h_{\mu\nu}\partial^\lambda 
h^{\mu\nu}+\partial_\mu h_{\nu\lambda}\partial^\nu h^{\mu\lambda}-\partial_\mu 
h^{\mu\nu}\partial_\nu h+\frac{1}{2}\partial_\lambda h\partial^\lambda h 
 -\frac{1}{2}m^2(h_{\mu \nu}h^{\mu \nu}-h^2) .
\ee
The relative sign in the mass term is quite essential for the theory to be consistent 
as a quantum theory because a ghost appears as a free particle without the tuning.
Hinterbichler \cite{Hinterbichler:2013eza}, pointed out that we may add new interaction terms 
to this model without generating any ghost. In four dimensions, there are two kinds of ghost-free interaction terms.
\begin{align}
\mathcal{L}_3 
\sim & \eta^{\mu_{1} \nu_{1} \mu_{2} \nu_{2} \mu_{3} \nu_{3}} h_{\mu_{1}\nu_{1}} h_{\mu_{2} 
\nu_{2}} h_{\mu_{3} \nu_{3}} \, ,
\label{nhhh} \\
\mathcal{L}_4
\sim & \eta^{\mu_{1} \nu_{1} \mu_{2} \nu_{2} \mu_{3} \nu_{3} \mu_{4} \nu_{4}} 
h_{\mu_{1} \nu_{1}} h_{\mu_{2} \nu_{2}} h_{\mu_{3} \nu_{3}} h_{\mu_{4} \nu_{4}} \, .
\label{nhhhh}
\end{align}
Here $\eta^{\mu_{1} \nu_{1} \cdots \mu_{n} \nu_{n}}$ is the product of $n$ 
$\eta_{\mu\nu}$ given by antisymmetrizing the indices $\nu_1$, $\nu_2$, $\cdots$, and 
$\nu_n$ 
Some examples are given by, 
\begin{align}
\label{h3c}
\eta^{\mu_{1} \nu_{1} \mu_{2} \nu_{2}} \equiv &  
\eta^{\mu_{1} \nu_{1}} \eta^{\mu_{2} \nu_{2}} - \eta^{\mu_{1} \nu_{2}} \eta^{\mu_{2} 
\nu_{1}}\, , \nn
\eta^{\mu_{1} \nu_{1} \mu_{2} \nu_{2} \mu_{3} \nu_{3}} \equiv & 
\eta^{\mu_{1} \nu_{1}}\eta^{\mu_{2} \nu_{2}} \eta^{\mu_{3} \nu_{3}} - \eta^{\mu_{1} 
\nu_{1}}\eta^{\mu_{2} \nu_{3}} \eta^{\mu_{3} \nu_{2}}
+ \eta^{\mu_{1} \nu_{2}}\eta^{\mu_{2} \nu_{3}} \eta^{\mu_{3} \nu_{1}} \nn
& - \eta^{\mu_{1} \nu_{2}}\eta^{\mu_{2} \nu_{1}} \eta^{\mu_{3} \nu_{3}}
+ \eta^{\mu_{1} \nu_{3}}\eta^{\mu_{2} \nu_{1}} \eta^{\mu_{3} \nu_{2}} - \eta^{\mu_{1} 
\nu_{3}}\eta^{\mu_{2} \nu_{2}} \eta^{\mu_{3} \nu_{1}} 
\, .
\end{align}
Using this notation, the Fierz-Pauli Lagrangian (\ref{FPLag}) is expressed as 
\be
\mathcal{L} 
= \frac{1}{2} \eta^{\mu_{1} \nu_{1} \mu_{2} \nu_{2} \mu_{3} \nu_{3}} 
\partial_{\mu_{1}} h_{\mu_{2} \nu_{2}} \partial_{\nu_{1}}  h_{\mu_{3} \nu_{3}}
 + \frac{m^2}{2} \eta^{\mu_{1} \nu_{1} \mu_{2} \nu_{2}} h_{\mu_{1} \nu_{1}} h_{\mu_{2} 
\nu_{2}} 
\ee 
In \cite{Ohara:2014vua}, it was proposed a new model of massive spin-two particles by 
adding the terms in (\ref{nhhh}) and (\ref{nhhhh}) to the Fierz-Pauli Lagrangian.

\begin{align}
\label{h3d}
\mathcal{L} 
= & \frac{1}{2} \eta^{\mu_{1} \nu_{1} \mu_{2} \nu_{2} \mu_{3} \nu_{3}} 
\partial_{\mu_{1}} h_{\mu_{2} \nu_{2}} \partial_{\nu_{1}}  h_{\mu_{3} \nu_{3}}
 + \frac{m^2}{2} \eta^{\mu_{1} \nu_{1} \mu_{2} \nu_{2}} h_{\mu_{1} \nu_{1}} h_{\mu_{2} 
\nu_{2}} \nn
& + \frac{\mu}{3!} \eta^{\mu_{1} \nu_{1} \mu_{2} \nu_{2} \mu_{3} \nu_{3}} 
h_{\mu_{1}\nu_{1}} h_{\mu_{2} \nu_{2}} h_{\mu_{3} \nu_{3}}
+\frac{\lambda}{4!} \eta^{\mu_{1} \nu_{1} \mu_{2} \nu_{2} \mu_{3} \nu_{3} \mu_{4} \nu_{4}} 
h_{\mu_{1} \nu_{1}} h_{\mu_{2} \nu_{2}} h_{\mu_{3} \nu_{3}} h_{\mu_{4} \nu_{4}} \nn
\end{align}
Here $m$, $\mu$ and $\lambda$ are parameters and the signs in front of $\mu$ and $\lambda$ are chosen to be opposite to them in \cite{Ohara:2014vua}.

Thanks to the ghost free property of the interactions, this theory does not contain any ghost and the particle description 
also holds in nontrivial vacua in some region of the parameter space spanned by $m^2$, $\lambda$, and $\mu$ \cite{Ohara:2014vua}. 

We should note that the model with cubic interactions, including the derivatives 
interactions, was first proposed in \cite{Folkerts:2011ev} before \cite{Hinterbichler:2013eza}.
%and it was also proved that there is no ghost in the model. 

\section{Global $U(1)$ theory}
\label{section3}
We can extend the model of massive spin-two theory by replacing the real field with the complex field. For the theory to be consistent with the 
global $U(1)$ symmetry, the cubic interaction is prohibited and the Lagrangian is given by
\begin{align}
\mathcal{L} = \eta^{\mu_1 \nu_1 \mu_2 \nu_2 \mu_3 \nu_3 } \partial_{\mu_1} h^{\dagger}_{\mu_2 \nu_2} \partial_{\nu_1} h_{\mu_3 \nu_3} 
+m^2 \eta^{\mu_1 \nu_1 \mu_2 \nu_2} h^{\dagger}_{\mu_1 \nu_1} h_{\mu_2 \nu_2} 
+ \frac{\lambda}{3!} \eta^{\mu_{1} \nu_{1} \mu_{2} \nu_{2} \mu_{3} \nu_{3} \mu_{4} \nu_{4}}
h^{\dagger}_{\mu_{1} \nu_{1}} h_{\mu_{2} \nu_{2}} h^{\dagger}_{\mu_{3} \nu_{3}} h_{\mu_{4} \nu_{4}}.
\label{A1}
\end{align}
The choice of $m^2>0$ guarantees stability around a trivial vacuum and absence of tachyonic state. Moreover, the theory does not have any nontrivial vacuum when $m^2$ and $\lambda$ are positive. 
Hence, $\lambda>0$ may suggest that the Hamiltonian is bounded 
from below in the analogy of the ordinary scalar field theory. 

The complex field $h_{\mu \nu}$ can be expressed as two real fields $a_{\mu \nu}$, $b_{\mu \nu}$
\be
h_{\mu \nu}=\frac{1}{\sqrt{2}} \left( a_{\mu \nu}+i b_{\mu \nu}\right).
\ee
Then, the action (\ref{A1}) describes an interacting real massive spin-two field theory. 

\begin{align}
\mathcal{L} &= \frac{1}{2} \eta^{\mu_1 \nu_1 \mu_2 \nu_2 \mu_3 \nu_3 } \partial_{\mu_1} a_{\mu_2 \nu_2} \partial_{\nu_1} a_{\mu_3 \nu_3} 
+\frac{m^2}{2} \eta^{\mu_1 \nu_1 \mu_2 \nu_2} a_{\mu_1 \nu_1} a_{\mu_2 \nu_2} + \frac{\lambda}{4!}\eta^{\mu_{1} \nu_{1} \mu_{2} \nu_{2} \mu_{3} \nu_{3} \mu_{4} \nu_{4}}
a_{\mu_{1} \nu_{1}} a_{\mu_{2} \nu_{2}} a_{\mu_{3} \nu_{3}} a_{\mu_{4} \nu_{4}}  \nonumber \\
&+\frac{1}{2} \eta^{\mu_1 \nu_1 \mu_2 \nu_2 \mu_3 \nu_3 } \partial_{\mu_1} b_{\mu_2 \nu_2} \partial_{\nu_1} b_{\mu_3 \nu_3} 
+\frac{m^2}{2} \eta^{\mu_1 \nu_1 \mu_2 \nu_2} b_{\mu_1 \nu_1} b_{\mu_2 \nu_2}+\frac{\lambda}{4!} \eta^{\mu_{1} \nu_{1} \mu_{2} \nu_{2} \mu_{3} \nu_{3} \mu_{4} \nu_{4}}
b_{\mu_{1} \nu_{1}} b_{\mu_{2} \nu_{2}} b_{\mu_{3} \nu_{3}} b_{\mu_{4} \nu_{4}}\nonumber \\
&+\frac{\lambda}{2 \cdot 3!}  \eta^{\mu_{1} \nu_{1} \mu_{2} \nu_{2} \mu_{3} \nu_{3} \mu_{4} \nu_{4}}
a_{\mu_{1} \nu_{1}} a_{\mu_{2} \nu_{2}} b_{\mu_{3} \nu_{3}} b_{\mu_{4} \nu_{4}}.
\label{A2}
\end{align}
The appearance of the interaction term between the $a$ and $b$ fields is the only nontrivial point, but it is easy to see that this theory is also ghost-free. 
The antisymmetric property of the $\eta$ symbol ensures that $a_{00}$ and $b_{00}$ appears linearly. In addition to this, there is no term containing $a_{0i} a_{00}$ 
, $a_{0i} b_{00}$, $b_{0i} a_{00}$ and $b_{0i} b_{00}$, which means that the equation of motion for the $0i$ component never gives a quadratic term in the $00$ component of the fields in the Lagrangian. 
Therefore, this system has the two constraints which are obtained by the variation of $a_{00}$ and $b_{00}$,

\begin{align}
&-\frac{\lambda}{3!} \eta^{i_1 j_1 i_2 j_2 i_3 j_3} a_{i_1 j_1} b_{i_2 j_2} b_{i_3 j_3} - \frac{\lambda}{3!} \eta^{i_1 j_1 i_2 j_2 i_3 j_3} a_{i_1 j_1} a_{i_2 j_2} a_{i_3 j_3}
-m^2 \eta^{i j} a_{ij}+ \eta^{i_1 j_1 i_2 j_2} \partial_{i_1}\partial_{j_1} a_{i_2 j_2} =0, \\
&-\frac{\lambda}{3!}  \eta^{i_1 j_1 i_2 j_2 i_3 j_3} b_{i_1 j_1} a_{i_2 j_2} a_{i_3 j_3} - \frac{\lambda}{3!} \eta^{i_1 j_1 i_2 j_2 i_3 j_3} b_{i_1 j_1} b_{i_2 j_2} b_{i_3 j_3}
-m^2 \eta^{i j} b_{ij}+ \eta^{i_1 j_1 i_2 j_2} \partial_{i_1}\partial_{j_1} b_{i_2 j_2} =0.
\end{align}
Here the Latin indices run from one to three.
Clearly, each equation specifies an independent hypersurface in the phase space and the $U(1)$ theory is really ghost-free. 

\section{Decoupling limit and Stability against quantum correction}
In this section, we study the behavior of the theory around the perturbative cutoff scale and the quantum stability. First, 
we introduce the Stuckelberg field.

\be
h_{\mu \nu} \rightarrow h_{\mu \nu} + \partial_{\mu} A_{\nu} + \partial_{\nu} A_{\mu} + 2 \partial_{\mu} \partial_{\nu} \phi.
\ee
After the diagonalizing the quadratic mixing terms between $h_{\mu \nu}$ and $\phi$ and canonically normalizing $\phi$, we find the most dangerous interactions for the perturbative unitarity,

\begin{align*}
& \sim \frac{\lambda}{m^6}  \eta^{\mu_1 \nu_1 \mu_2 \nu_2 \mu_3 \nu_3 \mu_4 \nu_4} h^{\dagger}_{\mu_1 \nu_1} \Pi_{\mu_2 \nu_2} \Pi_{\mu_3 \nu_3}^{\dagger} \Pi_{\mu_4 \nu_4}, \\
& \sim \frac{\lambda}{m^6}  \eta^{\mu_1 \nu_1 \mu_2 \nu_2 \mu_3 \nu_3 \mu_4 \nu_4} h_{\mu_1 \nu_1} \Pi_{\mu_2 \nu_2}^{\dagger} \Pi_{\mu_3 \nu_3} \Pi_{\mu_4 \nu_4}^{\dagger},  \\
& \sim \frac{\lambda}{m^6}  \eta^{\mu_1 \nu_1 \mu_2 \nu_2 \mu_3 \nu_3} \partial_{\mu_1} \phi^{\dagger} \partial_{\nu_1} \phi \Pi_{\mu_2 \nu_2}^{\dagger} \Pi_{\mu_3 \nu_3}. 
\end{align*}
Here we define $\Pi_{\mu \nu}$ as $ \partial_{\mu} \partial_{\nu} \phi$. The tree level amplitude for $\phi^{\dagger} \phi \rightarrow \phi^{\dagger} \phi$ or $h^{\dagger} \phi \rightarrow h^{\dagger} \phi$ scattering at energy $E$ goes as  $\mathcal{M} \sim \frac{\lambda E^6}{m^6}$. 
Thus, the theory becomes strongly coupled at the energy $E\sim m / \lambda^{\frac{1}{6}}$. We focus on the strongly coupled scale $\Lambda :=m / \lambda^{\frac{1}{6}}$ by taking the decoupling limit 
$m \rightarrow 0$, $\lambda \rightarrow 0$, while $\Lambda=m / \lambda^{\frac{1}{6}}$ is fixed. 

\begin{align}
\mathcal{L}= & \eta^{\mu_1 \nu_1 \mu_2 \nu_2 \mu_3 \nu_3} \partial_{\mu_1} h^{\dagger}_{\mu_2 \nu_2} \partial_{\nu_1} h_{\mu_3 \nu_3} + 2 \eta^{\mu_1 \nu_1 \mu_2 \nu_2} h^{\dagger}_{\mu_1 \nu_1} \Pi_{\mu_2 \nu_2}
+ 2 \eta^{\mu_1 \nu_1 \mu_2 \nu_2} h_{\mu_1 \nu_1} \Pi_{\mu_2 \nu_2}^{\dagger} \nonumber \\
& + \frac{16}{3!} \frac{1}{\Lambda^6}  \eta^{\mu_1 \nu_1 \mu_2 \nu_2 \mu_3 \nu_3 \mu_4 \nu_4} h^{\dagger}_{\mu_1 \nu_1} \Pi_{\mu_2 \nu_2} \Pi_{\mu_3 \nu_3}^{\dagger} \Pi_{\mu_4 \nu_4}
+ \frac{16}{3!} \frac{1}{\Lambda^6}  \eta^{\mu_1 \nu_1 \mu_2 \nu_2 \mu_3 \nu_3 \mu_4 \nu_4} h_{\mu_1 \nu_1} \Pi_{\mu_2 \nu_2}^{\dagger} \Pi_{\mu_3 \nu_3} \Pi_{\mu_4 \nu_4}^{\dagger}  
\end{align}
We diagonalize the quadratic term to obtain the kinetic term for the scalar field by redefining the field $h_{\mu \nu} \rightarrow h_{\mu \nu}+ \phi \eta_{\mu \nu}$.

\begin{align}
\label{A3}
\mathcal{L}= & \eta^{\mu_1 \nu_1 \mu_2 \nu_2 \mu_3 \nu_3} \partial_{\mu_1} h^{\dagger}_{\mu_2 \nu_2} \partial_{\nu_1} h_{\mu_3 \nu_3} + \frac{16}{3!} \frac{1}{\Lambda^6}  \eta^{\mu_1 \nu_1 \mu_2 \nu_2 \mu_3 \nu_3 \mu_4 \nu_4} h^{\dagger}_{\mu_1 \nu_1} \Pi_{\mu_2 \nu_2} \Pi_{\mu_3 \nu_3}^{\dagger} \Pi_{\mu_4 \nu_4}
\nonumber \\
& + \frac{16}{3!} \frac{1}{\Lambda^6}  \eta^{\mu_1 \nu_1 \mu_2 \nu_2 \mu_3 \nu_3 \mu_4 \nu_4} h_{\mu_1 \nu_1} \Pi_{\mu_2 \nu_2}^{\dagger} \Pi_{\mu_3 \nu_3} \Pi_{\mu_4 \nu_4}^{\dagger} -6\partial_{\mu} \phi^{\dagger} \partial^{\mu} \phi 
-\frac{32}{3!} \frac{1}{\Lambda^6}  \eta^{\mu_1 \nu_1 \mu_2 \nu_2 \mu_3 \nu_3} \partial_{\mu_1} \phi^{\dagger} \partial_{\nu_1} \phi \Pi_{\mu_2 \nu_2}^{\dagger} \Pi_{\mu_3 \nu_3},
\end{align}
The last term in (\ref{A3}) corresponds to the quartic charged Galileon term and the special structure of the interaction guarantees that quantum correction to the operators is absent \cite{de Rham:2013tfb}.
We can roughly estimate this correction as in \cite{ArkaniHamed:2002sp, Hinterbichler:2011tt}. Due to the nonrenormalization theorem, the induced operators 
which respect the Galilean symmetry has the following form,
\be
\frac{\partial^q(\partial^2 \phi)^p}{\Lambda^{3p+q-4}}.
\ee
Therefore, the relevant operator for the mass correction can be expected to take the form of $\frac{1}{\Lambda^2} (\partial \partial \phi)^2$. 
Then, considering the relation between $h$ and $\phi$, we find the correction is given by $\delta m^2 \sim \left(\frac{m^2}{\Lambda^2} \right) m^2  = \lambda^{1/3} m^2$
and the value of the mass is technically natural. On the other hand, the quantum effect might induce a ghost having a mass lower than the cutoff scale. 
When the general mass term of the massive spin-two field is given by the form of
\be
-\frac{1}{2} m^2 (h^{\mu \nu} h_{\mu \nu} -(1-a) h^2),
\ee
the the scale of the ghost mass $m_g$ is roughly estimated as $m^2_{g} \sim \frac{m^2}{a}$. Therefore, if the quantum correction breaks the Fierz-Pauli tuning, 
the ghost mass is comparable to the cutoff scale and this model is consistent as the effective field theory. Fortunately, the Fierz-Pauli tuning does not
break down at one loop level \cite{de Rham:2013tfc} in this model and the ghost mass is larger than $\Lambda$. The correction for the quartic potential term, however, seems to break the Galileon-type 
tuning and the scale is given as $\delta \lambda \sim \lambda^2 \left( \frac{\Lambda}{m} \right)^4 =\lambda^{1/3} \cdot \lambda$.
%, which also shows the technical naturalness.

\section{The behavior of the theory around vacua}
Next, we carry out the stability analysis of vacua of this theory as in \cite{Ohara:2015vua}. In the previous work, 
we found that the neutral massive spin-two field theory can have multiple stable vacua depending on the parameter contained in the theory. In this section, we 
are going to do the same analysis and clarify the difference between the charged theory and the neutral one. We mentioned in Sec. \ref{section3} the relation between the 
stability of the trivial vacuum and the parameters $m^2$ and $\lambda$: the both parameters have to take positive values for the model to be stable although it is unclear that 
the positiveness of $\lambda$ really make the Hamiltonian bounded from below [see discussion in \cite{Ohara:2015vua}].
Therefore, in this section, we concentrate on degenerate nontrivial vacua. For this purpose, consider the case of the mass parameter $m^2$ takes negative value, 
that is, $m^2 \rightarrow - |m^2|$. Then, the field acquires vacuum expectation value (VEV) and the system has nontrivial vacua where the particle description could hold. (Note that these nontrivial vacua do not correspond to the global
lowest energy of the system, because of the property of this model \cite{Ohara:2015vua}.) 
The nontrivial vacua are given by the following vacuum expectation value of $h_{\mu \nu}$,
\be
\label{VEV}
h^{\mathrm{VEV}}_{\mu \nu} = \frac{C \e^{i \theta}}{\sqrt{2}}  \eta_{\mu \nu} = \frac{1}{\sqrt{2}} \sqrt{\frac{3{|m^2|}}{\lambda}} \e^{i \theta} \eta_{\mu \nu}.
\ee
We obtain the Lagrangian in the broken phase by considering the fluctuation around the VEV.
\be
h_{\mu \nu}= h^{\mathrm{VEV}}_{\mu \nu} + H_{\mu \nu}
\ee
Then, the mass term takes the following form.
\begin{align}
{\mathcal L}_{\mathrm{mass}} =&-|m^2| \eta^{\mu_1 \nu_1 \mu_2 \nu_2} h^{\dagger}_{\mu_1 \nu_1} h_{\mu_2 \nu_2} \nonumber \\
=&-6 |m^2| C^2 -\frac{3}{\sqrt{2}}C |m^2| H -\frac{3}{\sqrt{2}}C |m^2| H^{\dagger} - |m^2| \eta^{\mu_1 \nu_1 \mu_2 \nu_2} H^{\dagger}_{\mu_1 \nu_1} H_{\mu_2 \nu_2} \nonumber 
\end{align}
%=&-\frac{|m^2|}{2} \eta^{\mu_1 \nu_1 \mu_2 \nu_2} a_{\mu_1 \nu_1} a_{\mu_2 \nu_2} -\frac{|m^2|}{2} \eta^{\mu_1 \nu_1 \mu_2 \nu_2} b_{\mu_1 \nu_1} b_{\mu_2 \nu_2} \nonumber \\
%&-6|m^2|C^2 - 3 |m^2|Ca
Here $H$ and $H^{\dagger}$ denote $\eta^{\mu \nu} H_{\mu \nu}$ and $\eta^{\mu \nu} H^{\dagger}_{\mu \nu}$ respectively.
The interaction term in the broken phase is also rewritten in terms of $H_{\mu \nu}$.
\begin{align}
{\mathcal L}_{\mathrm{int}} =& 3|m^2| C^2 + \frac{3}{\sqrt{2}}C|m^2| H+  \frac{3}{\sqrt{2}}C|m^2| H^{\dagger} +2 |m^2| \eta^{\mu_1 \nu_1 \mu_2 \nu_2} H^{\dagger}_{\mu_1 \nu_1} H_{\mu_2 \nu_2} \nonumber \\
+& \frac{|m^2|}{2} \eta^{\mu_1 \nu_1 \mu_2 \nu_2} H_{\mu_1 \nu_1} H_{\mu_2 \nu_2} + \frac{|m^2|}{2}  \eta^{\mu_1 \nu_1 \mu_2 \nu_2} H^{\dagger}_{\mu_1 \nu_1} H^{\dagger}_{\mu_2 \nu_2} + \sqrt{\frac{\lambda |m^2|}{6}} \eta^{\mu_1 \nu_1 \mu_2 \nu_2 \mu_3 \nu_3} H_{\mu_1 \nu_1} H^{\dagger}_{\mu_2 \nu_2} H_{\mu_3 \nu_3} \nonumber \\
+& \sqrt{\frac{\lambda |m^2|}{6}} \eta^{\mu_1 \nu_1 \mu_2 \nu_2 \mu_3 \nu_3} H^{\dagger}_{\mu_1 \nu_1} H_{\mu_2 \nu_2} H^{\dagger}_{\mu_3 \nu_3} + \frac{\lambda}{3!} \eta^{\mu_1 \nu_1 \mu_2 \nu_2 \mu_3 \nu_3 \mu_4 \nu_4} H^{\dagger}_{\mu_1 \nu_1} H_{\mu_2 \nu_2} H^{\dagger}_{\mu_3 \nu_3} H_{\mu_4 \nu_4} 
\end{align}
Therefore, the total Lagrangian is given by

\begin{align}
\label{A4}
{\mathcal L}_{\mathrm{BP}} =& \eta^{\mu_1 \nu_1 \mu_2 \nu_2 \mu_3 \nu_3 } \partial_{\mu_1} H^{\dagger}_{\mu_2 \nu_2} \partial_{\nu_1} H_{\mu_3 \nu_3} + |m^2| \eta^{\mu_1 \nu_1 \mu_2 \nu_2} H^{\dagger}_{\mu_1 \nu_1} H_{\mu_2 \nu_2} \nonumber \\
+& \frac{|m^2|}{2} \eta^{\mu_1 \nu_1 \mu_2 \nu_2} H_{\mu_1 \nu_1} H_{\mu_2 \nu_2}+ \frac{|m^2|}{2}  \eta^{\mu_1 \nu_1 \mu_2 \nu_2} H^{\dagger}_{\mu_1 \nu_1} H^{\dagger}_{\mu_2 \nu_2} + \sqrt{\frac{\lambda |m^2|}{6}} \eta^{\mu_1 \nu_1 \mu_2 \nu_2 \mu_3 \nu_3} H_{\mu_1 \nu_1} H^{\dagger}_{\mu_2 \nu_2} H_{\mu_3 \nu_3} \nonumber \\
+& \sqrt{\frac{\lambda |m^2|}{6}} \eta^{\mu_1 \nu_1 \mu_2 \nu_2 \mu_3 \nu_3} H^{\dagger}_{\mu_1 \nu_1} H_{\mu_2 \nu_2} H^{\dagger}_{\mu_3 \nu_3} + \frac{\lambda}{3!} \eta^{\mu_1 \nu_1 \mu_2 \nu_2 \mu_3 \nu_3 \mu_4 \nu_4} H^{\dagger}_{\mu_1 \nu_1} H_{\mu_2 \nu_2} H^{\dagger}_{\mu_3 \nu_3} H_{\mu_4 \nu_4}
\end{align}
The Lagrangian in the broken phase does not contain the Boulware Deser type ghost thanks to the antisymmetric tensor and is not $U(1)$ invariant.

Apparently, this looks that the system could contain one Nambu-Goldstone 
boson corresponding to the broken generator of $U(1)$ group. To verify this fact, let us focus on the quadratic part of this Lagrangian,
\begin{align}
{\mathcal L}^{(2)}_{\mathrm{BP}}&=\eta^{\mu_1 \nu_1 \mu_2 \nu_2 \mu_3 \nu_3 } \partial_{\mu_1} H^{\dagger}_{\mu_2 \nu_2} \partial_{\nu_1} H_{\mu_3 \nu_3} \nonumber \\
&+|m^2| \eta^{\mu_1 \nu_1 \mu_2 \nu_2} H^{\dagger}_{\mu_1 \nu_1} H_{\mu_2 \nu_2} + \frac{|m^2|}{2} \eta^{\mu_1 \nu_1 \mu_2 \nu_2} H_{\mu_1 \nu_1} H_{\mu_2 \nu_2}+ \frac{|m^2|}{2}  \eta^{\mu_1 \nu_1 \mu_2 \nu_2} H^{\dagger}_{\mu_1 \nu_1} H^{\dagger}_{\mu_2 \nu_2}.
\end{align}
The field $H_{\mu \nu}$ can be parametrized by two real fields $A_{\mu \nu}$ and $B_{\mu \nu}$.
\be
H_{\mu \nu}=\frac{1}{\sqrt{2}} \left(A_{\mu \nu} + i B_{\mu \nu} \right)
\ee
Then, we obtain 
\begin{align}
{\mathcal L}^{(2)}_{\mathrm{BP}}=& \eta^{\mu_1 \nu_1 \mu_2 \nu_2 \mu_3 \nu_3 } \partial_{\mu_1} H^{\dagger}_{\mu_2 \nu_2} \partial_{\nu_1} H_{\mu_3 \nu_3} \nonumber \\
&+|m^2| \eta^{\mu_1 \nu_1 \mu_2 \nu_2} H^{\dagger}_{\mu_1 \nu_1} H_{\mu_2 \nu_2} + \frac{|m^2|}{2} \eta^{\mu_1 \nu_1 \mu_2 \nu_2} H_{\mu_1 \nu_1} H_{\mu_2 \nu_2}+ \frac{|m^2|}{2}  \eta^{\mu_1 \nu_1 \mu_2 \nu_2} H^{\dagger}_{\mu_1 \nu_1} H^{\dagger}_{\mu_2 \nu_2} \nonumber \\
=& \frac{1}{2} \eta^{\mu_1 \nu_1 \mu_2 \nu_2 \mu_3 \nu_3 } \partial_{\mu_1} A_{\mu_2 \nu_2} \partial_{\nu_1} A_{\mu_3 \nu_3}+\frac{1}{2} m^2_A \eta^{\mu_1 \nu_1 \mu_2 \nu_2} A_{\mu_1 \nu_1} A_{\mu_2 \nu_2} + \frac{1}{2} \eta^{\mu_1 \nu_1 \mu_2 \nu_2 \mu_3 \nu_3 } \partial_{\mu_1} B_{\mu_2 \nu_2} \partial_{\nu_1} B_{\mu_3 \nu_3}
\label{B1}
\end{align}
where $m^2_A=2|m^2|$. 

According to Goldstone's theorem, the massless mode corresponds to the oscillation along the flat direction of the potential, which is given by the infinitesimal difference between two vacua: 
\be
\delta h^{\mathrm{VEV}}_{\mu \nu} = \frac{i \theta}{\sqrt{2}} C \eta_{\mu \nu}.
\ee
Therefore, the massless particle should be a scalar. On the other hand, the quadratic Lagrangian (\ref{B1}) shows that the system has one massive spin-two mode and one massless spin-two mode only.
From this fact, we find that the Nambu-Goldstone mode is absent because the massless field $B_{\mu \nu}$ is traceless and does not contain a scalar mode as far as we assume that perturbative description holds.
Furthermore, this Lagrangian has the nonderivative interactions for not only $A_{\mu \nu}$ but also $B_{\mu \nu}$ when we express (\ref{A4}) in terms of these two fields, which means that the degrees of freedom 
of the quadratic terms (\ref{B1}) do not coincide with the degrees of freedom of the full theory. 
\begin{align*}
\label{B2}
{\mathcal L}_{\mathrm{interactions}}
=&\sqrt{\frac{\lambda}{24} } m_A \eta^{\mu_1 \nu_1 \mu_2 \nu_2 \mu_3 \nu_3} A_{\mu_1 \nu_1} A_{\mu_2 \nu_2} A_{\mu_3 \nu_3}+\sqrt{\frac{\lambda}{24} } m_A \eta^{\mu_1 \nu_1 \mu_2 \nu_2 \mu_3 \nu_3} A_{\mu_1 \nu_1} B_{\mu_2 \nu_2} B_{\mu_3 \nu_3}\nonumber \\
&+\frac{\lambda}{4!}  \eta^{\mu_1 \nu_1 \mu_2 \nu_2 \mu_3 \nu_3 \mu_4 \nu_4} B_{\mu_1 \nu_1} B_{\mu_2 \nu_2} B_{\mu_3 \nu_3} B_{\mu_4 \nu_4} +\frac{\lambda}{4!}  \eta^{\mu_1 \nu_1 \mu_2 \nu_2 \mu_3 \nu_3 \mu_4 \nu_4} A_{\mu_1 \nu_1} A_{\mu_2 \nu_2} A_{\mu_3 \nu_3} A_{\mu_4 \nu_4} \nonumber \\
&+\frac{\lambda}{3! \cdot 2}  \eta^{\mu_1 \nu_1 \mu_2 \nu_2 \mu_3 \nu_3 \mu_4 \nu_4} A_{\mu_1 \nu_1} A_{\mu_2 \nu_2} B_{\mu_3 \nu_3} B_{\mu_4 \nu_4} 
\end{align*}

This fact strongly suggests that this model should be strongly coupled in the nontrivial vacua as in the discussion of \cite{Hinterbichler:2013eza} and the perturbative picture assumed in the above analysis should break down. 
This is the reason why the system seems not to have the Nambu-Goldstone mode: In the broken phase, the Nambu-Goldstone mode would exist, but the model does not have enough power to describe the dynamics of the massless 
scalar particle as an effective field theory. The explanation is perfectly consistent with the above observation that the Nambu-Goldstone mode is absent as far as the perturbative description is assumed.

As a result, the charged $U(1)$ theory cannot be defined around the nontrivial vacua but is defined only around the trivial vacuum, whose situation is different from the case of the neutral massive spin-two model (\ref{h3d}).

\begin{comment}
\begin{align}
{\mathcal L}_{\mathrm{potential}} =& |\lambda| \frac{|m^2|}{M^2}  C^4 + |\lambda| \frac{|m^2|}{M^2}  C^3 a +\frac{|\lambda|}{2} \frac{|m^2|}{M^2} C^2 \eta^{\mu_1 \nu_1 \mu_2 \nu_2} a_{\mu_1 \nu_1} a_{\mu_2 \nu_2}
+\frac{|\lambda|}{3!} \frac{|m^2|}{M^2} C \eta^{\mu_1 \nu_1 \mu_2 \nu_2 \mu_3 \nu_3} a_{\mu_1 \nu_1} a_{\mu_2 \nu_2} a_{\mu_3 \nu_3} \nonumber \\
&+ \frac{|\lambda|}{4!} \frac{|m^2|}{M^2} \eta^{\mu_1 \nu_1 \mu_2 \nu_2 \mu_3 \nu_3 \mu_4 \nu_4} a_{\mu_1 \nu_1} a_{\mu_2 \nu_2} a_{\mu_3 \nu_3} a_{\mu_4 \nu_4} 
+\frac{|\lambda|}{3!} \frac{|m^2|}{M^2} C^2 \eta^{\mu_1 \nu_1 \mu_2 \nu_2} b_{\mu_1 \nu_1} b_{\mu_2 \nu_2} \nonumber \\ 
&+\frac{|\lambda|}{4!} \frac{|m^2|}{M^2} \eta^{\mu_1 \nu_1 \mu_2 \nu_2 \mu_3 \nu_3 \mu_4 \nu_4} b_{\mu_1 \nu_1} b_{\mu_2 \nu_2} b_{\mu_3 \nu_3} b_{\mu_4 \nu_4} 
+\frac{|\lambda|}{3!} \frac{|m^2|}{M^2} C \eta^{\mu_1 \nu_1 \mu_2 \nu_2 \mu_3 \nu_3} a_{\mu_1 \nu_1} b_{\mu_2 \nu_2} b_{\mu_3 \nu_3} \nonumber \\
&+\frac{|\lambda|}{2\cdot3!} \frac{|m^2|}{M^2} \eta^{\mu_1 \nu_1 \mu_2 \nu_2 \mu_3 \nu_3 \mu_4 \nu_4} a_{\mu_1 \nu_1} a_{\mu_2 \nu_2} b_{\mu_3 \nu_3} b_{\mu_4 \nu_4}
\end{align}
\end{comment}

\section{Summary}
We have extended the new model of the massive spin-two field proposed in \cite{Ohara:2014vua} by imposing the global $U(1)$ symmetry and investigated 
several basic properties of the model. The difference from the neutral massive spin-two model is the existence of the interaction term between two kinds 
of fields, but this does not break the ghost-free property of the theory. The interaction may induce the quantum correction to the operators in the tree 
level Lagrangian and this could be also the cause for the ghost. Fortunately, the structure of the Lagrangian in the decoupling limit strongly suggests
the energy scale where the detune in the potential term happens can be made very high as far as the coupling constant $\lambda$ is very small. Based on this
discussion, we also study the property of vacua in this theory. While the new model of the neutral massive spin-two theory has nontrivial vacua if $m^2<0$ where the 
particle description holds, the charged theory can be only defined around the trivial vacuum because the degrees of freedom in the asymptotic region does not coincide with
the degrees of freedom of the full theory in the nontrivial vacua for any value of $m^2$ and $\lambda$.

In this paper, we have exclusively considered the global $U(1)$ theory, but it is interesting to extend the discussion to the $U(1)$ gauge theory where the massive spin-two 
particle is coupled with the photon. The local $U(1)$ symmetry is obtained by the replacement of the partial derivatives with covariant derivatives. The fact that the 
the charged massive spin-two theory only makes sense in the trivial vacuum suggests that the $U(1)$ gauge theory is also well behaved only in the trivial vacuum. Therefore, 
the Higgs mechanism might not be induced by the massive spin-two field in the model we proposed. Moreover, according to the preceding work
 by Porrati and  Rahman\cite{Porrati2008}, the cutoff scale of the perturbative unitarity is universal for the local $U(1)$ massive spin-two theory. This analysis, however, was done for 
the model which does not contain the self-interacting term for the spin-two particle. Hence, it is valuable to discuss the effect of the self-interaction to the cutoff scale
and we will study these subjects in the future.

\section*{Acknowledgments} 
I would like to thank Keisuke Harigaya, Takumi Kuwahara, Ryo Nagai, Taishi Ikeda and Professor Shin'ichi Nojiri for helpful discussions.
This research is supported by the Grant-in-Aid for JSPS Research Fellow No.16J04912.
\appendix

\begin{comment}

\section{Noether charge}
Let us define the infinitesimal transformation for the field as

\begin{align*}
h_{\mu \nu} & \rightarrow h_{\mu \nu}-i\alpha h_{\mu \nu} \
h^{\dagger}_{\mu \nu} & \rightarrow h^{\dagger}_{\mu \nu} + i \alpha h^{\dagger}_{\mu \nu}
\end{align*}
The usual procedure gives the Noether current for the theory:

\begin{align}
j^{\alpha} &= \frac{\partial \mathcal{L}}{\partial \left(\partial_{\mu} h_{\nu \rho} \right)} \delta h_{\nu \rho} + \frac{\partial \mathcal{L}}{\partial (\partial_{\mu} h^{\dagger}_{\nu \rho} )} \delta h^{\dagger}_{\nu \rho}  \nonumber \\
&= -i \eta^{\mu_1 \alpha \mu_2 \nu_2 \mu_3 \nu_3} \partial_{\mu_1} h^{\dagger}_{\mu_2 \nu_2}h_{\mu_3 \nu_3} + i \eta^{\alpha \nu_1 \mu_2 \nu_2 \mu_3 \nu_3} h^{\dagger}_{\mu_2 \nu_2} \partial_{\nu_1 } h_{\mu_3 \nu_3}
\end{align}
Therefore, the Noether charge is defined as $Q= \int d^3x j^0$.

\end{comment}

\end{document}